%% file: main.tex
\documentclass[journal]{IEEEtran}
\usepackage{graphicx}

\AtBeginDocument{%
  \providecommand\BibTeX{{%
    \normalfont B\kern-0.5em{\scshape i\kern-0.25em b}\kern-0.8em\TeX}}}

\usepackage{cite}
\usepackage[table]{xcolor}
\usepackage{amsmath,amsfonts}
\usepackage{algorithmic}
\usepackage{graphicx}
\usepackage{enumitem}
\usepackage{textcomp}
\usepackage{fancyhdr}
\usepackage{tikz}
\usepackage{multirow}
\usepackage{ulem}
\usepackage{comment}
\usepackage{pifont}
\usepackage{amssymb}

\newcommand{\cmark}{\ding{51}} %
\newcommand{\xmark}{\ding{55}} %
\usepackage{hyperref}  
\newcommand*{\rom}[1]{\expandafter\@slowromancap\romannumeral #1@}

\usepackage{orcidlink}
\begin{document}

\title{MATCH: Model-Aware TVM-based Compilation for Heterogeneous Edge Devices
}%

\newcommand\orcidicon[1]{\href{https://orcid.org/#1}{\mbox{\scalerel*{
\begin{tikzpicture}[yscale=-1,transform shape]
\pic{orcidlogo};
\end{tikzpicture}
}{|}}}}
\author{
\IEEEauthorblockN{ Mohamed Amine Hamdi \textsuperscript{1}
\orcidlink{0009-0006-0539-5007},
Francesco Daghero \textsuperscript{1}
\orcidlink{0000-0002-6215-8220},%
Giuseppe Maria Sarda\textsuperscript{2,3}
\orcidlink{0000-0001-6231-3553},
Josse Van Delm\textsuperscript{2}
\orcidlink{0000-0002-9503-403X}, \\%\orcidicon{0000-0002-9503-403X}, \\
Arne Symons\textsuperscript{2}
\orcidlink{0000-0002-4595-9384},
Luca Benini\textsuperscript{4}
\orcidlink{0000-0001-8068-3806}, %
Marian Verhelst\textsuperscript{2,3}
\orcidlink{0000-0003-3495-9263}, %
Daniele Jahier Pagliari\textsuperscript{1}
\orcidlink{0000-0002-2872-7071},
Alessio Burrello\textsuperscript{1,4}
\orcidlink{0000-0002-6215-8220}
}\\
\IEEEauthorblockA{
\textsuperscript{1}\textit{Politecnico di Torino}, Italy,
\textsuperscript{2}\textit{KU Leuven}, Belgium,
\textsuperscript{3}\textit{IMEC}, Belgium, 
\textsuperscript{4}\textit{University of Bologna}, Italy} {}
}
\maketitle

\input{Text/00_Abstract}

\maketitle

\input{Text/01_Intro}

\input{Text/02_Background}
\input{Text/03_Related}
\input{Text/04_1_MATCH}
\input{Text/04_2_Additional_target}
\input{Text/05_Results}
\input{Text/06_Conclusion}

\input{Text/Acknowledgments}
\bibliographystyle{IEEEtran}
\bibliography{references}

\end{document}

%% file: Text/00_Abstract.tex
\begin{abstract}
Streamlining the deployment of Deep Neural Networks (DNNs) on heterogeneous edge platforms, coupling within the same micro-controller unit (MCU) instruction processors and hardware accelerators for tensor computations, is becoming one of the crucial challenges of the TinyML field.
The best-performing DNN compilation toolchains are usually deeply customized for a single MCU family, and porting to a different heterogeneous MCU family implies labor-intensive re-development of almost the entire compiler. 
On the opposite side, retargetable toolchains, such as TVM, fail to exploit the capabilities of custom accelerators, resulting in the generation of general but unoptimized code.
To overcome this duality, we introduce MATCH, a novel TVM-based DNN deployment framework designed for easy agile retargeting across different MCU processors and accelerators, thanks to a customizable model-based hardware abstraction.
We show that a general and retargetable mapping framework enhanced with hardware cost models can compete with and even outperform custom toolchains on diverse targets while only needing the definition of an abstract hardware model and a SoC-specific API.
We tested MATCH on two state-of-the-art heterogeneous MCUs, GAP9 and DIANA. 
On the four DNN models of the MLPerf Tiny suite MATCH reduces inference latency by up to 60.88$\times$ on DIANA, compared to using the plain TVM, thanks to the exploitation of the on-board HW accelerator. Compared to HTVM, a fully customized toolchain for DIANA, we still reduce the latency by 16.94\%.
On GAP9, using the same benchmarks, we improve the latency by 2.15$\times$ compared to the dedicated DORY compiler, thanks to our heterogeneous DNN mapping approach that synergically exploits the DNN accelerator and the eight-cores cluster available on board.
\end{abstract}

%% file: Text/01_Intro.tex
\section{Introduction}
\noindent The increasing focus on efficiently executing Deep Neural Networks (DNNs) on low-power MCUs has led to the proliferation of domain-specific Systems-On-Chip (SoCs)~\cite{gap8,gap9,genc2021gemmini, ueyoshi2022diana, jain2022tinyvers}.
Three crucial features characterize these architectures: first, being designed for the specific task of accelerating DNNs at the extreme edge, they are usually \textit{OS-less}, in order to minimize both the memory requirements for the software stack and the associated performance overheads. Second, they are usually \textit{heterogeneous platforms} composed of general-purpose cores in charge of simple control tasks (e.g., receiving input data and sending DNN outputs) and a set of accelerators optimized for DNN workloads. As a last point, they commonly avoid HW caches in favor of SW scratchpad memories to reduce the required silicon area, exploiting the regularity of memory accesses in DNN workloads.
Therefore, fully exploiting the hardware resources of these platforms when executing a DNN is a non-trivial task, which requires in-depth knowledge of the SoC and its memory hierarchy.

Automated deployment toolchains address these challenges by generating target-dependent code from high-level DNN representations (e.g., Python, ONNX, etc), abstracting hardware details from users~\cite{chen2018tvm,burrello2021dory}. However, existing toolchains for OS-less edge devices are usually either \textit{overly generic} or \textit{overly specific}. Generic toolchains, such as the popular TVM~\cite{chen2018tvm} or the MLIR-based TinyIREE~\cite{tinyiree2022}, have extensive support for diverse DNN operators, but only generate code for common CPU and GPU targets, given that they do not have access to in-depth specifics of the target hardware and, therefore, are not able to exploit SoC-specific features such as HW accelerators, ISA extensions, memory hierarchy dimensions, etc. 
Autotuning engines~\cite{zheng2020ansor} can partially address this issue by iteratively generating different versions of the code,  profiling it, and using search strategies such as reinforcement learning or genetic algorithms to converge to an optimized implementation.
On the other hand, they result in very long compile times for large search spaces~\cite{htvm} and usually require the availability of hardware-in-the-loop for profiling.

On the other side of the spectrum, target-specific toolchains provide top performance on specific SoCs, fully exploiting the target's accelerators and memory hierarchy. However, their support for DNN operators is limited, and adding a new one requires an in-depth knowledge of both the HW target and the full toolchain. Moreover, they are commonly monolithic software stacks that embed hardcoded hardware-dependent heuristics for memory management, operation scheduling, and tiling~\cite{burrello2021dory}.
Consequently, adding support for a new DNN accelerator or a new SoC requires a labor-intensive overhaul of the toolchain or, in the worst case, the creation of an almost entirely new toolchain, leading to large delivery delays of the final product.

A seminal work that combines these two worlds is HTVM~\cite{htvm},
which extends TVM with a customizable plug-in for hardware accelerator support~\cite{burrello2021dory}.
The extension is responsible for the generation of optimized code for operations that can be accelerated on the target SoC's accelerator, while TVM can be used to deploy non-accelerable kernels on the main CPU.
However, HTVM still relies on a monolithic tool (DORY), in which hardware-specific cost information is deeply intertwined with the code base, thus suffering from the above-mentioned limitations in terms of extensibility to new targets.

In this paper, we describe a step in the direction of solving this \textit{generic toolchains customization problem} proposing \textbf{MATCH}, a \textbf{M}odel-\textbf{A}ware \textbf{T}VM-based \textbf{C}ompiler for \textbf{H}eterogeneous edge devices. We focus specifically on the automatic optimization of tiling and loop ordering, two key steps to maximize throughput and energy efficiency of DNNs layers, which require strongly hardware-dependent decisions (e.g., on the optimal tile size and dimension)~\cite{burrello2021dory,htvm}.
MATCH plugs a DNN Design Space Exploration (DSE) tool in TVM, and to support a new SoC, it requires: i) a high-level HW model to let the DSE tool generate optimized layer-specific scheduling of loops and ii) SoC-specific APIs for the different accelerators, to manage and trigger DNN kernel execution.
Our goal is to demonstrate that MATCH can deploy DNNs as efficiently as hardware-specific toolchains while requiring significantly fewer changes to support new hardware.
In our experiments, a single compiler engineer, not involved in designing the hardware platforms, has been able to add support for a new SoC in MATCH in less than 1 week, including several refinement iterations for the cost model.

In summary, the main contributions of this work are the following:
\begin{itemize}[leftmargin=*]
    \item We propose MATCH, a new compiler that extends the TVM compilation flow with a DSE tool for DNN layer scheduling. Although the methodology is in principle orthogonal to the specific DSE engine, this work describes an implementation based on ZigZag~\cite{zigzag}, an open-source tool that identifies optimal temporal mappings, exploiting an abstraction of the workload and target hardware.
    To exploit HW heterogeneity, we propose a pattern-matching mechanism that exploits ZigZag's modeling to match each DNN operator with the best hardware module to execute it.
    \item As ZigZag only yields the temporal mapping, we enhance it with (i) an input interface to read DNN layers workloads from TVM,  (ii) a set of easily-modifiable APIs to support new hardware, and (iii) a novel code generation step. 
    \item We benchmark MATCH on two different heterogeneous MCUs: GAP9~\cite{gap9}, featuring an 8 RISC-V cores cluster and a flexible DNN HW accelerator, and DIANA~\cite{ueyoshi2022diana}, featuring two DNN accelerators working at different bitwidths. On a wide set of Convolutional Neural Network (CNN) layers, MATCH reduces the average latency by 119.08$\times$ on GAP9 and by 83.18$\times$ on DIANA, compared to the plain TVM solution. %
    \item On end-to-end DNN networks from the MLPerf Tiny benchmark~\cite{mlcommons}, MATCH achieves similar performance compared to the best SoCs-specific open-source toolchains, with 2.15$\times$ and 16.94\% lower average latency on GAP9 and DIANA, respectively. It also outperforms TVM on both platforms by 67.83$\times$ and 60.88$\times$.
\end{itemize}
We open-source our code at: \texttt{\url{https://github.com/eml-eda/match}}.
The rest of the manuscript is organized as follows:
Sec.~\ref{sec:background} and Sec.~\ref{sec:related} discuss the required background and related works respectively.
Sec.~\ref{sec:methods} describes MATCH's core components. In Sec.~\ref{sec:extension}, we show how to add support for a new SoC in MATCH.
Sec.~\ref{sec:results} reports the experimental results, and Sec.~\ref{sec:conclusion} concludes the paper.

%% file: Text/02_Background.tex
\section{Background}
\label{sec:background}
\noindent The optimized execution of DNNs at the edge is a complex task that can benefit from both software and hardware optimizations.
On the software side, different solutions, such as neural architecture search, pruning, or quantization, have been proposed to reduce DNNs' complexity while preserving accuracy \cite{kuzmin2024pruning, wang2019haq, white2023neural, coexpnas, lightnas, s3nas}. On the hardware side, many solutions have been proposed for accelerating DNN workloads, resulting in a plethora of complex heterogeneous SoCs \cite{gap8,gap9, ueyoshi2022diana, MAX78000}. To fill the gap between high-level DNN models and such complex hardware targets, \textit{AI compilation} frameworks \cite{
zheng2020ansor, roller, burrello2021dory, chen2018tvm, htvm, nntool, CubeAI, tinyiree2022, tflitemicro2021, dnnvm} are used to generate target-specific code that efficiently exploits the available hardware resources. 

Our effort focuses on defining a new AI compiler, MATCH, that not only fulfills its scope of porting DNNs architecture on heterogeneous edge devices but also provides a lightweight interface for compiler engineers that simultaneously allows them to achieve three seemingly contrasting goals: i) \textit{deployment efficiency}, i.e., close-to-optimal performance, by carefully orchestrating the available HW resources; ii) \textit{broad support} for existing and upcoming DNN operators; iii) \textit{easy extensibility} to future hardware targets.
In the rest of this section, we first give an overview of existing heterogeneous hardware platforms for DNN execution at the edge and then provide the required background on AI compiler technology.

\subsection{Edge Platforms for DNN Inference}
The execution of DNNs in typical general-purpose OS-less edge devices, such as Microcontrollers (MCUs), is not an easy task due to the requirements these devices need to deliver for an acceptable quality of service, e.g. performance, within a limited power envelope.
On the other hand, DNN inference is a highly parallel workload, prevalently composed of operations such as convolutions and General Matrix Multiplications (GEMMs), opening the possibility of aggressive hardware optimization.
In this direction, modern SoCs for edge DNN inference are becoming increasingly \textit{heterogeneous}; they usually include one or more DNN accelerators, usually built as an array of multiply-and-accumulate (MAC) units to parallelize tensor processing workloads (i.e. convolutions, matrix multiplication, etc), coupled with general-purpose cores to execute less compute-intensive layers and a multi-level. Cores and accelerators typically exchange data through a multi-level, software-managed memory hierarchy to minimize DNNs tensors access time.

Several examples of heterogeneous MCUs are available both in academia~\cite{hp_ssc2020, ima_ssc2022, ueyoshi2022diana} and industry~\cite{MAX78000,gap9, gap8}, with embedded accelerators that vary from multi-cores Single Instruction Multiple Data (SIMD) / Single Instruction Multiple Threads (SIMT) units to systolic arrays.
One notable academic example is the Parallel Ultra-Low Power (PULP) family of devices~\cite{garofalo2020pulp}, which is built around a main control core that dispatches digital signal processing (DSP) tasks to a cluster of 8 identical RISC-V cores with specific ISA extensions such as SIMD MACs and memory operations with pointer post-increment.
A more aggressively heterogeneous MCU, DIANA, introduced in~\cite{ueyoshi2022diana}, couples a RISC-V core used for IO-control and workload dispatching with two highly specialized DNN accelerators, a digital SIMD array, and an Analog In-Memory Computing (AIMC) one.
In industry, the MAX78000~\cite{MAX78000} by Analog Devices enhances an ARM Cortex-M4 MCU with a CNN accelerator based on a fixed systolic array structure. 
GAP9, by GreenWaves, is an industrial embodiment of PULP, that couples a main RISC-V controller core with a flexible DNN accelerator (Called NE16) as well as a general-purpose 8-cores cluster~\cite{gap9}.
In this work, we target two of these heterogeneous devices to demonstrate the generality of our AI compiler, namely DIANA and GAP9.
In Section~\ref{sec:extension}, we demonstrate the flexibility of MATCH in supporting these two strongly different HW targets, achieved simply by customizing HW models and DNN operators APIs.

\subsection{AI compilers}
\begin{figure}[t]
    \centering
    \includegraphics[width=0.9\columnwidth]{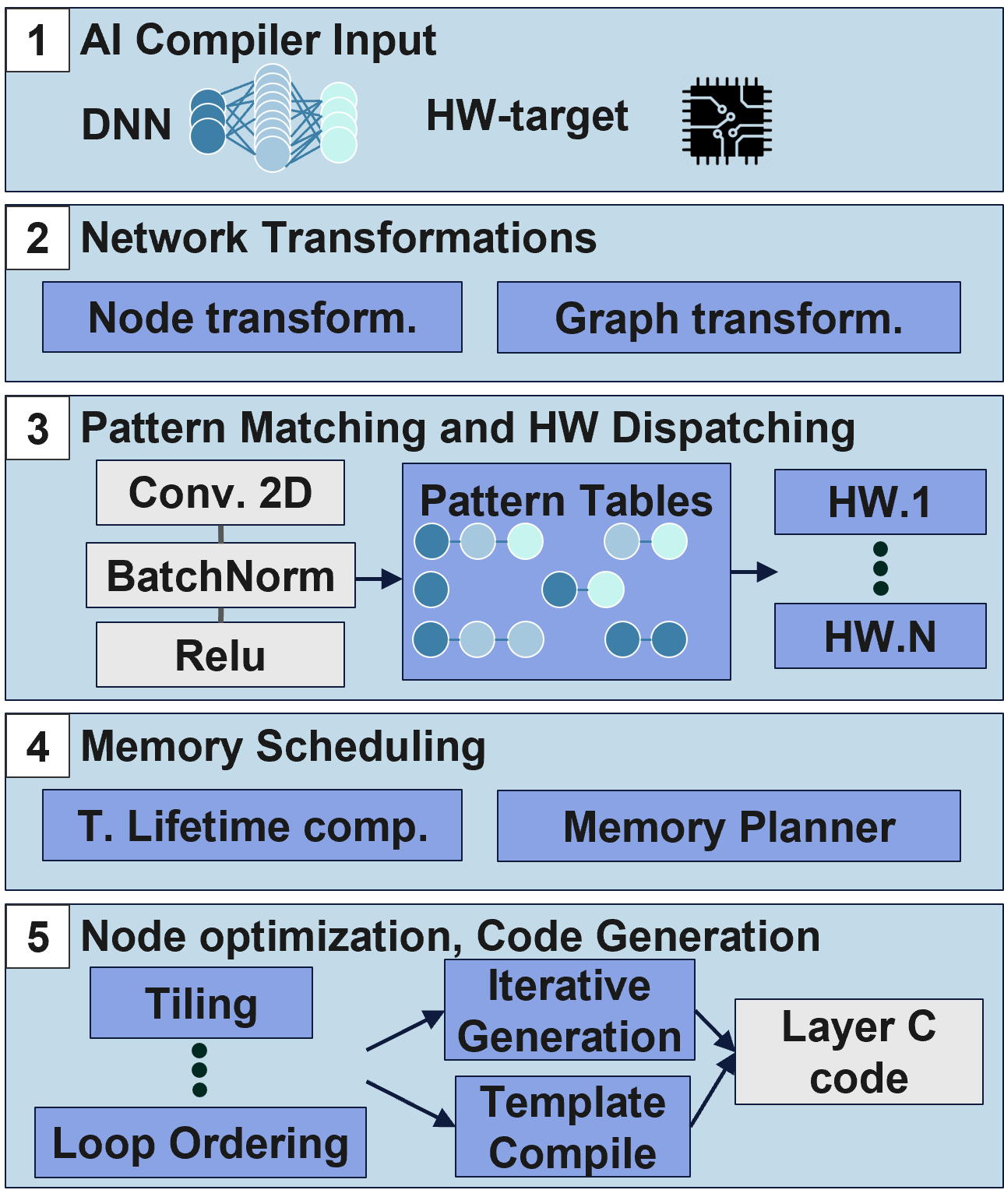}
    \caption{Overview of the components that are part of an AI compiler.}
    \label{fig:aicompiler}
\end{figure}
The innovations brought by researchers and vendors on both new HW platforms and DNN models are often bottlenecked by the lack of a sufficiently flexible yet performant middleware layer that allows the execution of new models on newly developed HW. Recently, many works have focused on this layer of the stack, usually referred to as the \textit{AI compiler}. In the context of an OS-less heterogeneous SoC, an AI Compiler consists of a toolchain to translate HW-unaware DNN formats, such as ONNX or PyTorch models, into HW-specific compiled code for their optimized execution into CPUs, GPUs, and dedicated accelerators. These tools are particularly important in edge computing, where HW architectures are highly heterogeneous with tight memory constraints.
Fig.\ref{fig:aicompiler} depicts the main AI compiler steps, detailed below.
\subsubsection{Network Transformations}
the compilation of AI models typically starts from a computational graph representing the workload, where nodes represent operations over tensors, abstracted as edges. Network transformations can be categorized into HW-aware and HW-unaware ones. HW-unaware passes include optimizations such as operator fusion.
On the other hand, HW-aware transformations consider the target hardware's characteristics to optimize the computational graph further, enhancing performance and efficiency, for instance, by replacing operations with equivalent ones that are better supported by the hardware, or adapting data layouts and quantization formats.

\subsubsection{Pattern Matching and HW Dispatching}  after applying transformations to the network's computational graph, the AI compiler offloads the execution of each node of the optimized graph to a specific hardware module. This step maps high-level operations in the computational graph to low-level hardware-specific operations or kernels. Pattern matching is applied on the graph to identify nodes that are supported by each HW module, to fully leverage SoC-specific capabilities, such as specialized instructions or acceleration units.

\subsubsection{Memory Scheduling} the last graph-level pass is usually memory scheduling, whose goal is minimizing the peak memory occupation for activations based on tensor lifetimes. Effective memory scheduling ensures that data is efficiently allocated and deallocated, reducing memory overhead and making the model execution feasible on devices with limited volatile memory space.

\subsubsection{Operator-Specific Optimizations} Finally, operator-specific optimizations are applied to each subgraph matched in step 2, to enhance the performance of individual operations onto the hardware module they have been assigned to. One crucial optimization technique is \textit{loop tiling}, which partitions activation and weights tensors into smaller, more manageable blocks or tiles. This enables more efficient data access patterns, particularly by maximizing data reuse within faster memory levels. Ideally, combining memory tiling with \textit{loop ordering} can enable reaching a performance similar to the one obtained with a theoretically infinite last-level cache. We will call the specific combination of loop tiling and ordering for a layer the \textit{layer schedule}.

After applying all steps, the final code is generated; here, either a template-based approach \cite{burrello2021dory, htvm, zheng2020ansor, CubeAI, nntool} or a lowering code-generation \cite{tinyiree2022, chen2018tvm, tflitemicro2021} pipeline can be used. 
Depending on the strategy, the kernel's code (i.e., the innermost convolution/GEMM operation, after the application of fusion, tiling, loop ordering, etc.) is imported from HW-specific hand-tuned libraries with optimized performance \cite{burrello2021dory} or produced together with the full layer code through auto-tuning pipelines with HW-in-the-loop \cite{zheng2020ansor}.

%% file: Text/03_Related.tex
\section{Related Works}
\label{sec:related}
\input{tables/deployment-frameworks-comparison}
\noindent Table \ref{tab:mcucompilers} summarizes the main state-of-the-art AI compilers from academia and industry chronologically. We categorize them into three groups based on the following criteria: i) the ease of extending them to a new HW target, ii) the ease of adding a new DNN operator, and iii) the performance achieved.

A first group of toolchains is designed to be highly extensible but often at the cost of generating unoptimized HW-unaware code. For example, TensorFlow Lite for Microcontrollers \cite{tflitemicro2021} is a popular framework that allows the conversion of TensorFlow models into optimized C++ kernels paired with a minimal runtime. To improve performance, TFLite allows the plugging of hand-written backends. However, it lacks advanced optimization passes, such as loop ordering and tiling, which are essential to fully exploit hardware capabilities. 
Two of the most popular AI compilers that support edge devices are Apache TVM \cite{chen2018tvm} and the Multi-level intermediate representation (MLIR)-based Tiny Intermediate Representation Execution Environment (TinyIREE)~\cite{tinyiree2022}.
The first generates bare-metal code for several general-purpose targets, including CPUs, GPUs, and MCUs. Adding new targets or new SW operators is straightforward, given that operators are expressed at a high level, using their algebraic function, without considering HW-specific features or possible optimizations such as ISA extension exploitation, tensor reuse, or data stationarity.
Similarly, TinyIREE proposes a generic lowering toolchain that exploits the MLIR paradigm. While offering a flexible and modular progressive lowering infrastructure, it does not natively embed passes for tiling, loop reordering, and other HW-dependent optimizations.

To cope with the lack of specialization of these compilers, HW vendors designed tools that exploit HW-specific information to improve utilization and, therefore, performance. CubeAI \cite{CubeAI} focuses on optimizing computation and memory usage for STM32 MCUs, allocating tensors in different DRAM regions, and exploiting different CMISIS-NN-based \cite{lai2018cmsis} kernels for different layer geometries. However, both the kernels and the code generation are tailored exclusively for this family of MCUs, not allowing them to be extended to other HW targets. NNTool, the Greenwaves deployment framework for GAP8 \cite{gap8} and its successor, GAP9, outperforms all the competitors in terms of energy consumption on the four MLPerf Tiny benchmarks. DORY \cite{burrello2021dory} reaches comparable performance on the same HW with an open-source tool based on the PULP-NN kernel library \cite{garofalo2020pulp}. Both tools exploit knowledge about the memory hierarchy to optimally tile the layer and orchestrate the kernel execution partitioning each tensor appropriately and moving it between the main memory and a SW-managed last-level cache. DORY formalizes tensor tiling as a constraint programming problem, embedding target-specific heuristics in its code.
All these tools share common difficulties in supporting upcoming DNN layers and topologies, given that they are constructed monolithically, with corner cases on specific layer geometries, network topologies, or individual operators hard-wired within multiple compiler passes.

Recognizing the need for a balance between extensibility and performance, some tools have begun to address this dichotomy.
Ansor \cite{zheng2020ansor} exploits TVM and an auto-tuning mechanism to optimize DNN layer scheduling but requires extensive exploration of compilation parameters (e.g., tiling factors), with HW-in-the-loop, making it time-consuming and often impractical for custom DNN hardware accelerators. Specifically, each operator is iteratively compiled with different sets of parameters and benchmarked on the hardware to find the optimal configuration~\cite{moreau2019hardware}.
While this auto-tuning mechanism is effective, executing all possible scheduling and tiling alternatives requires exploring thousands of variants for each kernel. Running them on the target is time-consuming. Therefore, the exploration space is often pruned, leading to sub-optimal final solutions. Furthermore, custom DNN  hardware accelerators are currently not natively supported.
Another example is ROLLER \cite{roller}, which employs a deterministic algorithm for optimal loop order and splitting search. The best deployment configuration is found through a Look-Up Table (LUT)-based approach, that is, an offline profiling of microkernels with pre-defined layer geometries is used to construct a LUT of layer performances. However, its extensibility is limited, given that the LUT should contain every possible layer hyperparameter alternative for HW targets characterized by a non-linear dependency between tensor geometry and latency (i.e., most real-world devices), leading to unfeasible compilation times.

The most promising approach is currently HTVM \cite{htvm}, a framework that integrates DORY into TVM, aiming to combine the best of both worlds by offloading DNN layers that can be accelerated to DORY.  The tool exploits the Bring Your Own Codegen (BYOC) infrastructure~\cite{chen2021bring} of TVM, which allows offloading pre-defined patterns of the network's computational graph to external user-defined routines optimized for specific targets such as accelerators. While this approach improves flexibility maintaining very high performance, it retains the limitations of DORY regarding extensibility to new hardware.

Given the limitations of existing tools, there is a clear need for an AI compiler that combines ease of extensibility with high performance. MATCH aims to fulfill this need by providing a framework that is: i) flexible and easily extensible to both new HW targets, by only providing limited HW-specific information, and to new unsupported operators, by exploiting TVM general routines; ii) capable of generating optimized code for diverse hardware targets, maximizing the utilization of the on-board compute units and accelerators.

%% file: tables/deployment-frameworks-comparison.tex
\begin{table*}[!ht]
    \centering
    \renewcommand{\arraystretch}{1.3}
    \caption{State-of-the-art in AI compilation for edge devices.}
    \begin{tabular}{|l|l|l|l|l|l|l|l|}
    \hline
        \textbf{Name} & \textbf{Open} & \textbf{Supported} & \textbf{Cost} & \textbf{Search} & \textbf{Platform} & \textbf{Operator} & \textbf{Perf.} \\
        & \textbf{Source} & \textbf{Targets} & \textbf{Model} & \textbf{Algorithm} & \textbf{Extension} & \textbf{Extension} &    \textbf{Achieved} \\ \hline
        \textbf{TVM,2018} \cite{chen2018tvm} & \cmark & CPU, GPU, MCU & None & None & \cellcolor{green} & \cellcolor{green} & \cellcolor{red} \\ \hline
        \textbf{Ansor,2020} \cite{zheng2020ansor} & \cmark & CPU, GPU, MCU & HW-in-the-loop & Reinforcement Learning & \cellcolor{yellow} & \cellcolor{green} & \cellcolor{yellow} \\ \hline
        \textbf{TFLite,2020} \cite{tflitemicro2021} & \cmark & MCUs & None & None &  \cellcolor{green} & \cellcolor{green} & \cellcolor{red} \\ \hline
        \textbf{DORY,2020} \cite{burrello2021dory} & \cmark & MCU, Accelerators & Coarse-grain Analytical & Linear Constraint Progr & \cellcolor{red} & \cellcolor{red} & \cellcolor{green} \\ \hline
        \textbf{NNTool,2020} \cite{nntool} & \xmark & GAP8, GAP9 & Fine-grain Analytical & Rules-driven & \cellcolor{red} & \cellcolor{red} & \cellcolor{green} \\ \hline
        \textbf{ROLLER,2022} \cite{roller} & \cmark & CPU, GPU & HW-based LUT  & Model-driven & \cellcolor{yellow} & \cellcolor{yellow} & \cellcolor{green} \\ \hline
        \textbf{CubeAI,2022} \cite{CubeAI} & \xmark & STM32 MCUs & None & None & \cellcolor{red} & \cellcolor{red} & \cellcolor{green} \\ \hline
        \textbf{TinyIREE,2022} \cite{tinyiree2022} & \cmark & CPU, GPU, MCU & None & None & \cellcolor{green} & \cellcolor{green} & \cellcolor{red} \\ \hline
        \textbf{HTVM,2023} \cite{htvm} & \cmark & MCU, Accelerators & Coarse-grain Analytical & Linear Constraint Progr & \cellcolor{red} & \cellcolor{green} & \cellcolor{green} \\ \hline
        \textbf{MATCH, Our} & \cmark & Heterogeneous MCUs & Fine-grain Analytical & Genetic-based & \cellcolor{green} & \cellcolor{green} & \cellcolor{green} \\ \hline
    \end{tabular}
    \label{tab:mcucompilers}
\end{table*}

%% file: Text/04_1_MATCH.tex
\section{MATCH}
\label{sec:methods}
\noindent To address the dualism between flexible-yet-inefficient and optimized-yet-inflexible toolchains, we introduce MATCH, a novel framework that enhances TVM with HW-aware deployment capabilities using the BYOC framework. Starting from a Python-level DNN model, MATCH generates optimized HW-specific C code to deploy the DNN on OS-less heterogeneous devices. MATCH exploits TVM-specific features, i.e., the BYOC interface and the Patter Matcher, to be able to simultaneously deploy optimized operators with external user codebases, while falling back to a default unoptimized solution for \textit{un-matched} operators. MATCH is written in Python $>$3.8.

\begin{figure}[t]
    \centering
    \includegraphics[width=\columnwidth]{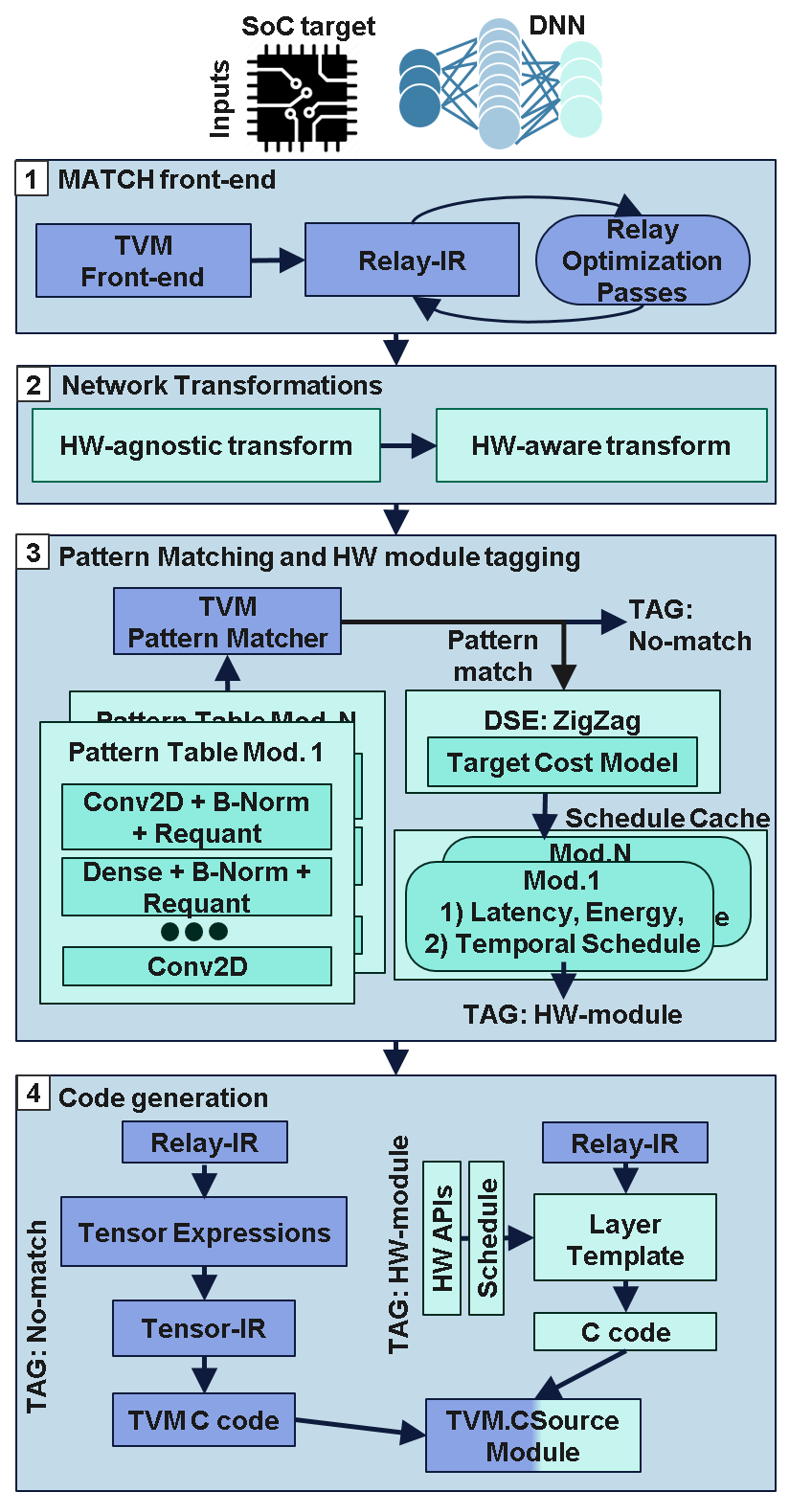}
    \caption{MATCH flow. TVM default components are colored in dark blue.} \label{fig:flow}
\end{figure}

Fig.~\ref{fig:flow} shows a high-level overview of the deployment flow using MATCH, which is composed of four main stages.
In the rest of this section, we describe the Framework Frontend and Network Transformations in Sec.~\ref{sec:match_front_end}, the Pattern Matching and HW-aware Dispatching in Sec~\ref{sec:match_pattern_matcher}, and the Code Generation in Sec~\ref{sec:match_codegen}.
In Sec.~\ref{sec:extension}, we further detail how to add a new HW target to MATCH, with two specific examples: DIANA and GAP9.

MATCH currently targets Convolutional Neural Networks (CNNs), since these models are very popular in extreme-edge applications. Its extension to transformers is planned as future work as these newer ML models are  just starting to be targeted for tinyML scenarios~\cite{tinytimeseriestrans}.
This, and other extensions, will be facilitated by the modularity of our approach.
Hereinafter, we use the following notation for the hyperparameters of a 2D Convolutional layer: IX/IY/C for the horizontal, vertical, and channel dimensions of the input tensor; OX/OY/K for the respective output dimensions; FX/FY for the spatial dimensions of the convolutional weights kernel.

\subsection{Framework Front-End and Network Transformations}
\label{sec:match_front_end}
MATCH receives two inputs to generate the target-specific C code: the NN model and the HW target definition. The target definition encompasses a list of \textbf{hardware execution modules}, i.e., units that can be used to execute one or more DNN operators.
Each hardware execution module contains the memory description (i.e., hierarchy, dimensions, and connections), a list of offloadable patterns, the set of specific APIs used, and a cost model for each operator. In Sec. \ref{sec:extension}, we detail all the information needed to support a new target.

The first step executed by the compiler is the NN model reading: we rely on the TVM frontend, which accepts a wide range of formats, from quantized Keras models to ONNX.
The input format is translated to Relay, the intermediate representation used internally by TVM to describe and manipulate the model graph. 

After this first step, standard HW-agnostic transformations are applied to the network nodes using the TVM \textit{ExprVisitor}, such as dead node elimination and constants folding. Starting from the intermediate representation generated after the HW-agnostic transformations, the graph is further manipulated by \textbf{HW-aware} transformations, i.e., ones that consider HW-specific features such as the number of processing elements (PE) or the presence of specific HW components.
HW-aware transformations can belong to two families: the first one encompasses patterns rewriting, which exploits the TVM  \textit{DFPatternCallback} to match a pattern in the graph and transform it into a new one. The second family, similarly to the HW-agnostic passes, exploits the TVM \textit{ExprVisitor} operator to traverse the full graph and apply transformations to individual nodes, such as adding parameters, changing data type or layout, to allow the mapping of operators to specific accelerators.
\input{tables/network_transformations}

As an example, we reported in Tab.~\ref{tab:net_transformations} the transformations applied when deploying a network to the GAP9 platform: first, constants are folded, and dead nodes are eliminated; then, integerization is applied to transform all tensors to int8 data type. A layout transformation is applied to store all activation tensors in NHWC data format, as required by the backend kernel libraries that we adopt for this target, namely PULP-NN \cite{garofalo2020pulp} for the RISC-V cluster, and a custom library for the NE16 accelerator\footnote{In the NHWC layout for activation tensors, two nearby memory locations store pixels relative to the same spatial position, but successive channels (C), as opposed to NCHW, the default layout in PyTorch, in which two contiguous cells store pixels relative to the same channel and two neighboring spatial locations along the horizontal dimension (W)~\cite{lai2018cmsis, garofalo2020pulp}.}.
Then, each mul-add-div sequence, used for re-quantizing a layer's output, is transformed into a re-quantization node implementing the arithmetic function $f(x) = (x*M+B)>>S$, removing the div operator, which is not supported by the NE16 and slow on the cluster. For the NE16 accelerator, padding is further applied to match the input tensor dimensions to the accelerator's spatial parallelism, and weights are re-ordered to match the custom data layout needed.
Note that the HW-agnostic transformations can be selected from a large pool of pre-defined passes, while each different HW clearly has its own set of HW-aware passes.

\subsection{Pattern Matching and Accelerator-aware Dispatching}
\label{sec:match_pattern_matcher}
After applying the network graph transformation pipeline, MATCH assigns the execution of each part of the DNN to a selected HW module through its Pattern Matcher module.
The Pattern Matcher analyzes the intermediate NN graph and tags groups of nodes (patterns) with a reference to the HW module that will execute them.
More specifically, MATCH can associate to each pattern either an HW module or a \textit{not-matched} tag; in the latter case, the TVM default code generation pipeline is invoked, and the layer will be executed on the main CPU of the SoC. If an HW module supporting the layer is found, MATCH exploits a DSE tool to generate an optimized schedule for it.
MATCH builds on top of the TVM's internal pattern matching system to facilitate the definition of new SoC-specific patterns.
Each HW module definition includes a Pattern Table that lists all patterns that can be offloaded to it. In detail, each pattern comprises the set of nodes to be matched, the replacement (one or more Relay nodes), and an additional Pattern Constraint that defines a set of rules that must be fulfilled for the match to be valid. 
These rules can verify if the input/output layouts, the quantization formats, the layer hyper-parameters (e.g., the convolutional kernel dimensions), etc, in the candidate graph pattern are compatible with those supported by the HW module.

Note that compared to other State-of-the-Art (SoA) AI compilers, such as DORY or HTVM, which \textit{separately} support accelerators and MCUs but do not allow the orchestration of multiple HW modules on the same SoC, MATCH implements an iterative exploration algorithm to offload patterns to the HW execution unit with the minimum expected latency/energy among those supporting a given pattern, thus enabling the combined usage of multiple HW modules \textit{on the same SoC}.
To do so, MATCH sequentially explores the Pattern Tables of each individual HW module and invokes the DSE tool for each positive match. For each combination of (pattern, node parameters, HW module), the DSE tool provides a twofold output: the best schedule and its predicted latency/energy. If multiple HW modules match a pattern, MATCH selects the one that minimizes one of these metrics and tags the pattern accordingly.
To cope with patterns contained within others (e.g., Conv-only versus Conv + Batch Norm + ReLU), we heuristically select the largest one, assuming that node fusion is always convenient to reduce latency and energy.

\subsubsection{Model-based DSE Engine}
As a DSE tool, MATCH employs the optimizer included in ZigZag~\cite{zigzag}, which can search both for the optimal \textit{spatial mapping} of a DNN layer onto a PE array (optimizing the number of PEs and their parallelism axes) and for the optimal \textit{temporal mapping} or \textit{schedule} (i.e., operator's loop ordering and loop tiling). Currently, in MATCH, we do not explore spatial mapping and focus only on temporal mapping optimization. Indeed, the spatial mapping is a fixed input, since we target already manufactured accelerators and hardware devices with pre-existing backend kernels with known parallelization strategies.
To optimize the temporal mapping, we parse the provided workload definition and then explore the search space using the LOMA~\cite{loma} engine. The latter generates all valid and non-equivalent scheduling candidates using the Loop Prime Factors approach~\cite{loma}. 
It then allocates each loop's operands to the lowest non-full memory level. An analytical hardware performance cost model guides the selection of the optimal schedule among the generated ones.
Importantly, LOMA supports \textit{uneven} mapping, i.e., with different tensors tiled in different memory levels.
We extend ZigZag and LOMA to take into account double-buffering. Further, we extend them to support a more general accelerator model, including systolic arrays and more generic multi-core accelerators.

Notice that to support a new hardware target, users of MATCH do not need to modify the DSE tool's internals; they just need to provide the corresponding operators' cost models.
A detailed example of how these analytical cost models can be easily and quickly defined is provided in Sec. \ref{sec:extension}.

\subsection{Code Generation}
\label{sec:match_codegen}
MATCH includes two code generation ``branches". For un-matched graph patterns, the \textit{fall-back codegen branch} is taken, and MATCH generates C code exploiting one of TVM's default targets. The NN node in Relay IR is lowered to Tensor Expressions and then into TIR, a lower-level IR that allows representing information such as loop unrolling and layer tiling. Then, TVM produces generic C code for pre-defined targets such as x86/ARM/RISC-V CPUs. Note that during this lowering step, TVM is unaware of any specific ISA instructions or dedicated HW blocks.

On the other hand, if a pattern has been tagged with a specific HW module, the \textit{specialized codegen branch} is executed. To generate the optimized C code, we provide a generic \textit{Layer Template}, which is compiled with the i) pattern hyperparameters, ii) the spatial/temporal scheduling provided by the DSE, and iii) platform-specific APIs. We use the Python Mako template library for this step.

To generate the network's main file, MATCH uses the default TVM generator that receives the layer function prototypes and the tensor dimensions to allocate the memory. We used TVM's default hill-climbing algorithm for memory allocation.

\subsubsection{Layer Template Compilation}
\begin{figure}[t]
    \centering
    \includegraphics[width=\columnwidth]{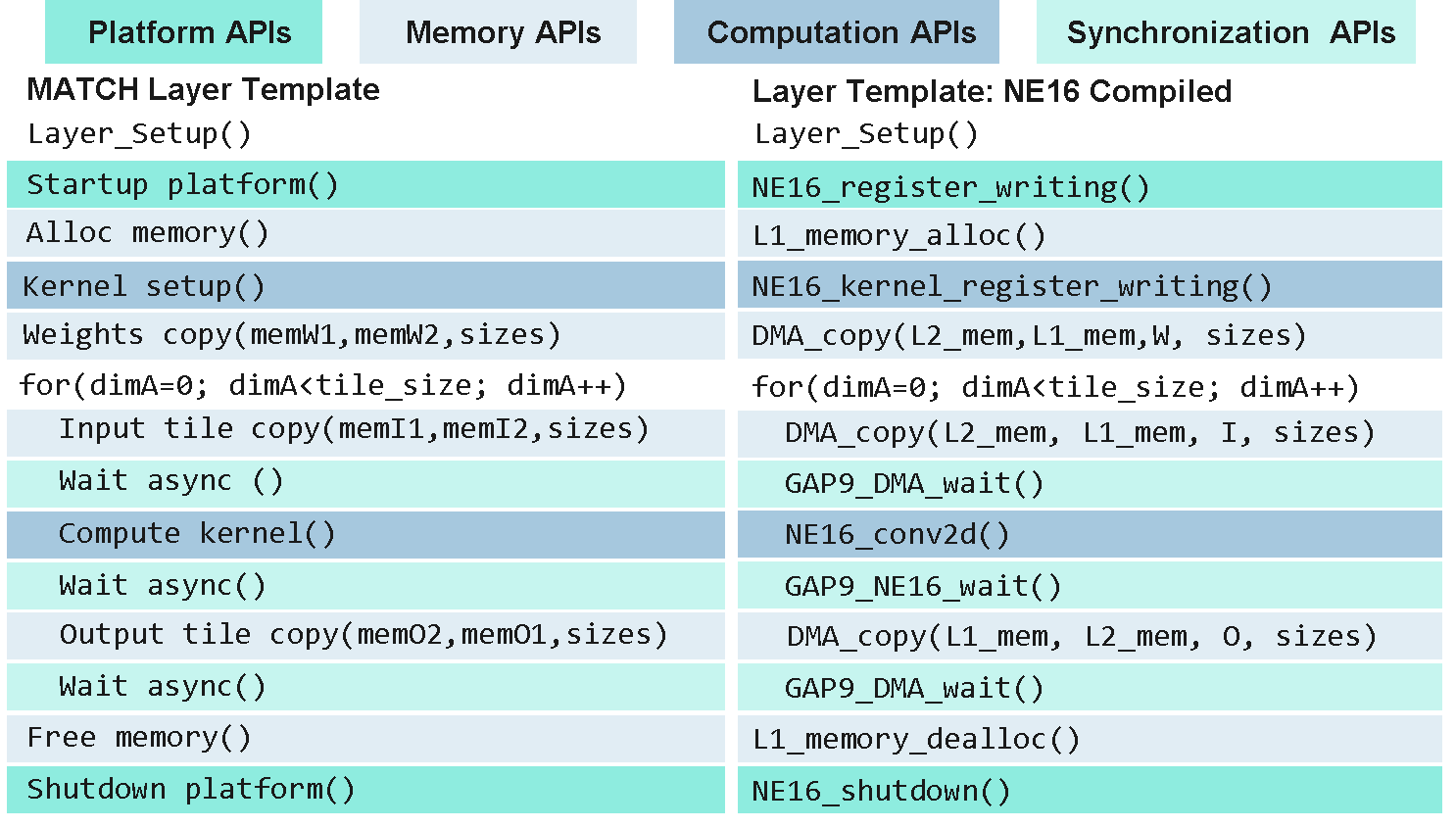}
    \caption{Overview of the layer template and its conversion to target-specific code.}
    \label{fig:template}
\end{figure}
First, default patterns hyperparameters such as layer geometry, padding, or stride are added to the template. Then, the cached DSE spatio-temporal scheduling is used to define the layer loops order, the loop splitting, and the tensor tiles dimensions. The memory transfers are also correctly placed to minimize the latency overhead, following the DSE scheduling and supporting both single- and double-buffering, depending on the provided DSE output. Finally, the pattern's computation kernel (i.e., a wrapper to the target's backend library) is inserted in the innermost loop, where all the input data have already been transferred to the innermost memory level. 
In the code generated using the template, all function calls are still platform-agnostic: for instance, a memory-copy does not call the SoC-specific DMA function, but a generic \texttt{MATCH\_memory\_copy}.

In the last code generation step, MATCH embeds HW-specific user-defined APIs in the template. These APIs are defined in the HW model for each target, by specifying their name in an API object.
We categorized APIs in 4 families:
\begin{itemize}
\item{\textbf{Platform APIs}}: they are used to allocate, deallocate, or configure the corresponding HW modules. For instance, for an HW accelerator, these APIs could be used to write on memory-mapped registers in order to configure the module for executing a specific operation.
\item{\textbf{Memory APIs}}: these APIs facilitate memory management by enabling the allocation and deallocation of different memory levels, handling data transfers between them, for instance utilizing DMA calls for efficiency, and computing pointer offsets when employing specific data layouts.
\item{\textbf{Synchronization APIs}}: MATCH allows for asynchronous and synchronous data management.
Therefore, it exposes APIs to synchronize after memory transfers, kernel computation, or both. Thanks to these APIs, we provide the possibility to perform both single- and double-buffered memory transfers.
\item{\textbf{Computational APIs}}: these are the backend library calls mentioned above. They receive the whole pattern context to set up and execute the correct computational kernel with its corresponding hyperparameters.
\end{itemize}

Fig. \ref{fig:template} reports the layer template on the left and the corresponding compilation with GAP9 NE16 APIs for a convolutional kernel on the right.

%% file: tables/network_transformations.tex
\begin{table}[!ht]
    \centering
    \caption{HW-agnostic and Hw-aware graph transformations applied in MATCH when deploying a network to GAP9.}
    \begin{tabular}{|l|l|l|}
    \hline
        \textbf{Network Transformations} & \textbf{Description} & \textbf{Type} \\ \hline
        \multicolumn{3}{l}{\textbf{HW-agnostic}}\\ \hline
        \multirow{2}{3cm}{\textbf{Dead Node Removal and Constant Folding}} & \multirow{2}{3cm}{Merging of contiguous dead/constant nodes} & \multirow{2}{1cm}{Graph Visitor} \\
         &  &  \\ \hline
        \textbf{Integerization} &  Quantize operations & Graph \\
        & and weights & Visitor \\ \hline
        \textbf{Layout transformation} & Switch between  & Graph \\
        & NCHW/NHWC & Visitor \\ \hline
        \multicolumn{3}{l}{\textbf{HW-aware}}\\ \hline
        \textbf{Requant sequence} & Transform division into  & Pattern \\
        & a right shift operation & Matching \\ \hline
        \textbf{Padding and slicing} & Padding and slicing & Graph \\
        & before/after matched sequence & Visitor \\ \hline
        \multirow{2}{3cm}{\textbf{Weights Transformation}} & \multirow{2}{3cm}{Storage of weights in custom data layout} & \multirow{2}{1cm}{Graph Visitor} \\
         &  &  \\ \hline
    \end{tabular}
    \label{tab:net_transformations}
\end{table}

%% file: Text/04_2_Additional_target.tex
\section{Adding a new MATCH HW Target}\label{sec:extension}
\begin{figure}[t]
    \centering
    \includegraphics[width=\columnwidth]{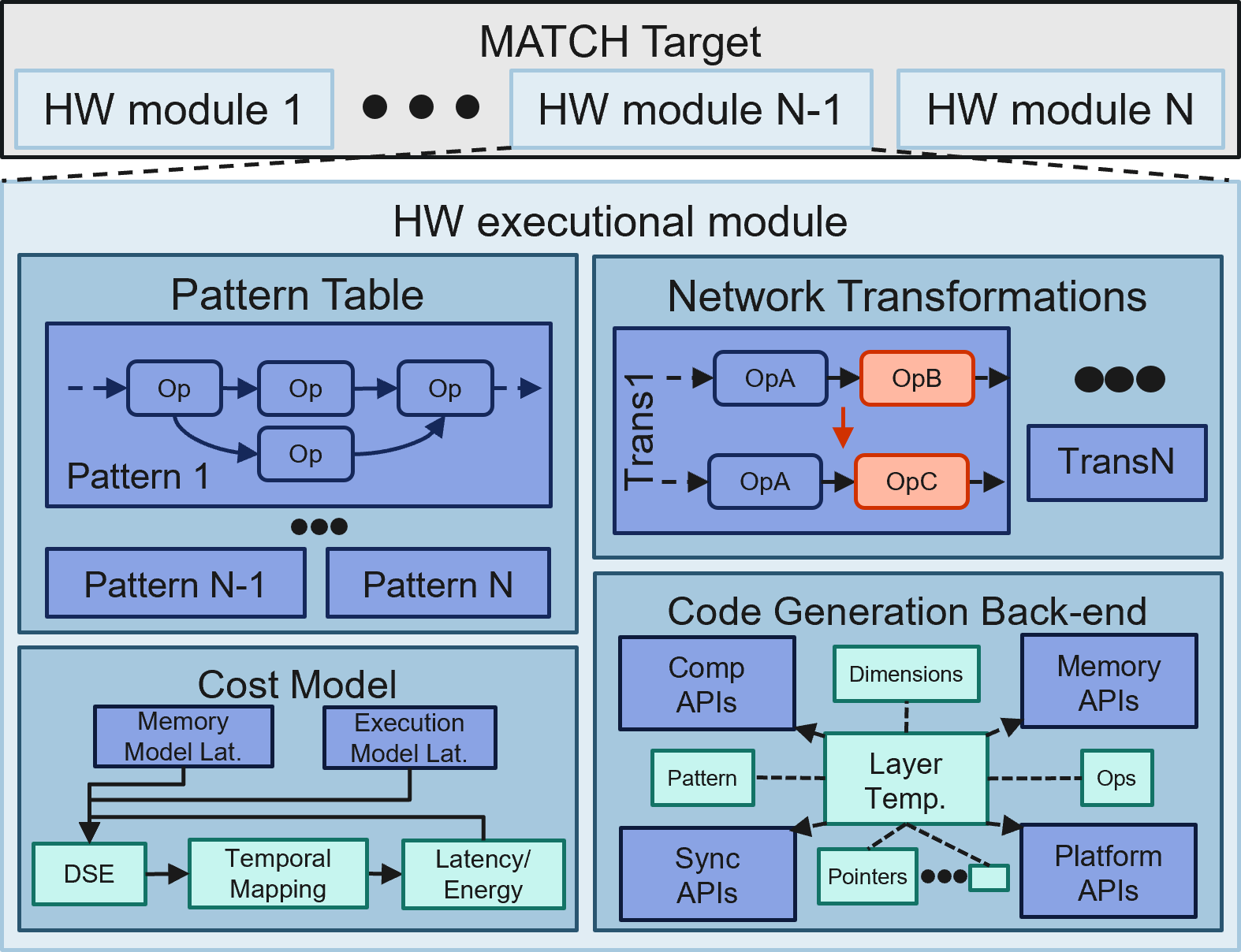}
    \caption{Example of a generic MatchTarget, detailing how an HW execution module is composed.}
    \label{fig:HW_modules}
\end{figure}
\noindent To extend MATCH to a new HW target, we provide the \textbf{MatchTarget} class, which can encompass one or more \textit{HW Execution Modules}. Each HW Execution Module contains four key components, as shown in Figure 4.

First, the above-mentioned \textbf{Pattern Table} lists the supported patterns for the module. Patterns define specific sequences of operations that the hardware module can efficiently execute. For instance, a pattern might include a 2D convolution followed by a batch normalization and a re-quantization step. 

The \textbf{Cost Model} is used for generating the correct schedule for each supported operator pattern during the DSE process. 
The cost model is defined analytically as a generic Python function taking information on the matched pattern, including its composing nodes, their hyper-parameters, data layouts, quantization formats, etc, and returning a scalar (e.g., the pattern's latency or energy). MATCH can support models with different levels of detail, ranging from rough to almost-cycle-accurate estimates. Those can take into account information such as memory bandwidth, HW spatial parallelism, memory hierarchy, ISA extensions, etc.
The most important property of a cost model is \textit{rank preservation} between schedules, i.e., if the estimated latency for schedule A is $<$ than the one for schedule B, then the same should hold for the actual on-device latencies.
Further, as previously mentioned, notice that the cost model is crucial to discern between multiple \texttt{MatchTarget} instances that support the same operations, if available, and select the most efficient one. Therefore, rank preservation across different HW modules is also crucial.

A set of \textbf{Network Transformations} is defined to be applied to the neural network both before and after graph partitioning. As mentioned in Sec. \ref{sec:match_front_end}, these transformations manipulate the graph for the specific hardware, modifying properties such as input tensor layout or replacing individual operations with equivalent ones, if the HW Module expects them in a peculiar way.

Finally, a \textbf{Code Generation Backend} is present for each HW module: this includes the specialization of the above-mentioned APIs for the target hardware, which replace the generic MATCH APIs in the final generated C code. Notice that APIs can be either HW module-specific or SoC-specific, providing a flexible and extensible framework for supporting new hardware targets.

Below, we provide two examples of HW targets, DIANA (only the digital accelerator) and GAP9, including its cluster of 8 general-purpose cores and its AI accelerator, NE16.

\subsection{DIANA Model Customization}
\begin{figure}[t]
    \centering
    \includegraphics[width=.98\columnwidth]{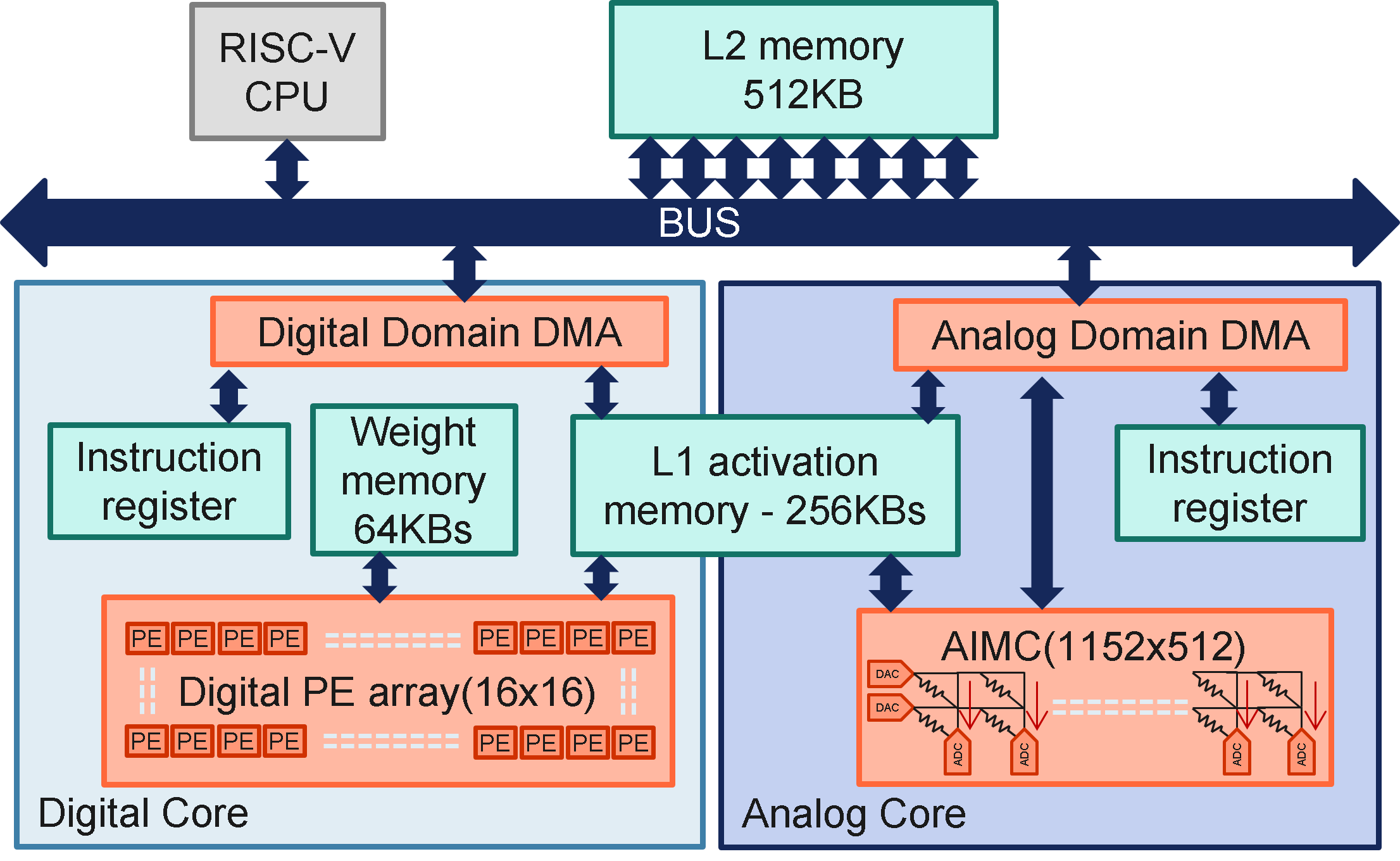}
    \caption{DIANA architecture. We report the computational units in orange and the memory units in light green.}
    \label{fig:DIANA}
\end{figure}
The first platform we target is DIANA~\cite{ueyoshi2022diana}, shown in Fig.~\ref{fig:DIANA}.
It features a RISC-V MCU, and two accelerators: a digital 8-bit accelerator and an Analog-in-Memory Computing (AIMC) ternary one.
Our work targets only the digital one since we only target 8-bit integer networks. However, the analog accelerator can be supported as an additional HW Execution Module of the DIANA \texttt{MatchTarget}.
The digital accelerator is a 2D SIMD array of 16×16 PEs, that can achieve up to 256 8-bit MAC operations per cycle, allowing execution of convolutions / fully-connected layers with re-quantization, ReLU, and pooling operations directly at the output. 
The DIANA accelerator implements 2D Convolutions by spatially unrolling the output channels (K) and output spatial width (OX) in the two physical dimensions of the array.
Instead, fully connected (FC) layers are spatially unrolled along input and output neurons.
We stored the corresponding patterns in the module's Pattern Table.
The accelerator has a 256kB L1 activation memory and a 64kB private weight memory, while the SoC L2 main memory is 512kB.
Transfers between different memory levels are handled through DMA blocking calls. In order to correctly and fully utilize the PE array, DIANA layer primitives require spatially unrolled dimensions (K and OX) to be multiple of 16 and a custom NCHW-derived data layout. 
Both these requirements are handled through Network Transformation passes, exploiting spatial padding and static weights reshapes (not adding overhead at runtime).

The ZigZag DIANA performance Cost Model includes two main contributions $L_{ops}$ and $L_{mem,i,j}$, respectively, the latency of inner loop computations in L1 and the latency for memory transfers between the $i$-th and $j$-th hierarchy levels. $L_{ops}$ is derived automatically from the spatial parallelism information provided as input, coupled with coefficients that account for the number of computation cycles in L1. For DIANA, this corresponds to the latency for reading inputs, performing MACs, and writing outputs (1 cycle each), plus the application of elementwise operators to the outputs, and their storage into the activation memory (23 cycles).
$L_{mem,1,2}$ models the memory transfer cycles and the performance overheads caused by L2-to-L1 DMA calls.
This component is derived from the memory bandwidth information and the transferred tensors dimensions. Additionally, the shape of the transfers influences the performance overhead, which we established to be 70-cycles for each chunk of data stored contiguously in memory. If a data block is not stored contiguously, the overhead is multiplied by the number of contiguous sub-blocks.
Since DIANA transfers data synchronously, ZigZag computes the overall latency as $L = L_{ops} + L_{mem,1,2}$.

The DIANA code generator includes the specific HW accelerator invocation and DMA APIs. No Synchronization APIs are needed, given the blocking nature of the DMA and the presence of a single active accelerator.

\begin{figure}[t]
    \centering
    \includegraphics[width=.98\columnwidth]{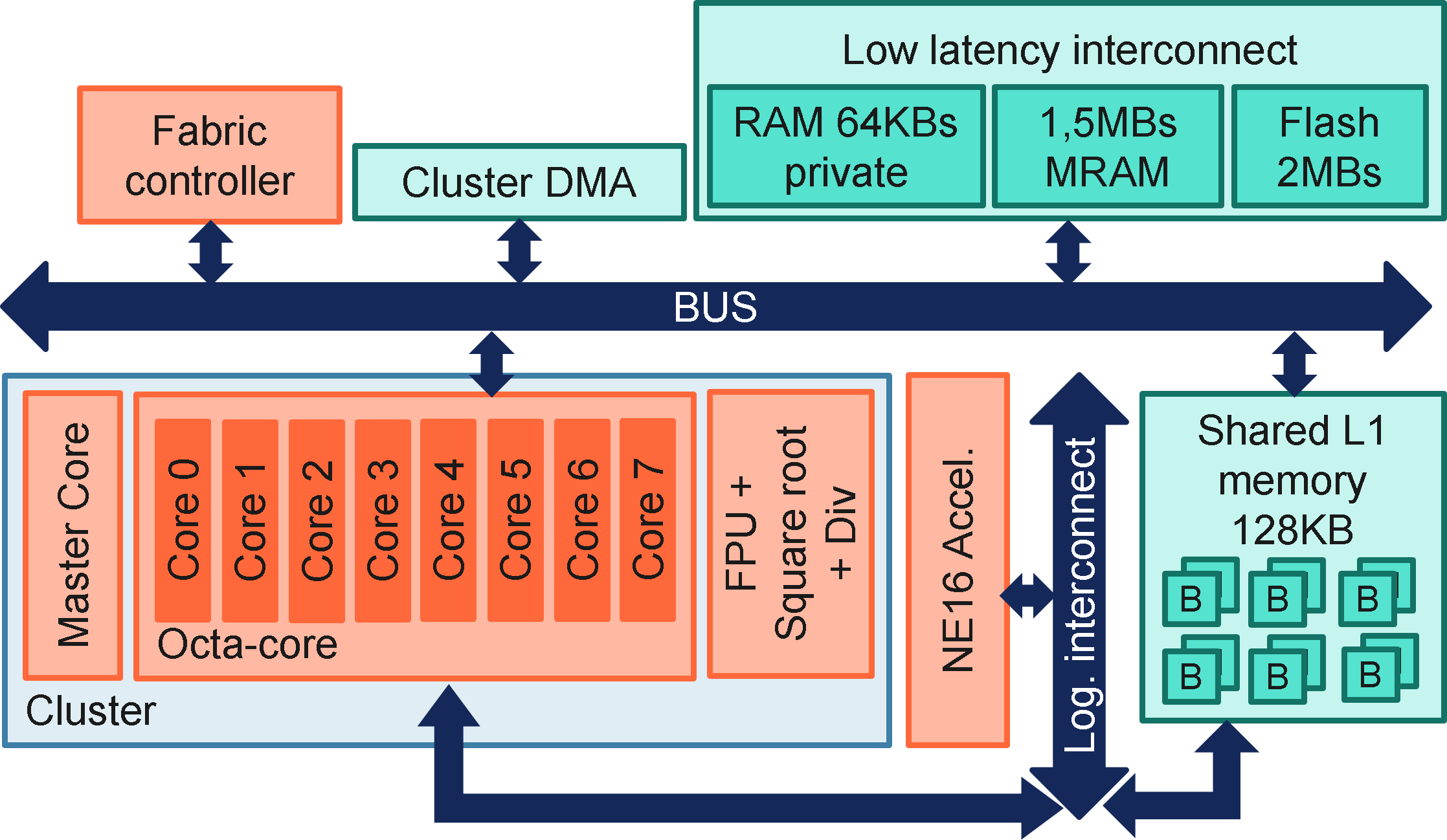}
    \caption{GAP9 architecture. We report the computational units in orange and the memory units in light green.}
    \label{fig:GAP9}
\end{figure}

\subsection{GAP9 Model Customization}\label{sec:gap9model}
GAP9 combines a control MCU with a programmable multi-core compute cluster and a dedicated AI accelerator, NE16 (Figure \ref{fig:GAP9}).
All cores use a RISC-V ISA enhanced with custom DSP-oriented extensions. Noteworthy, GAP9's \texttt{MatchTarget} is composed of two different HW Execution Modules (cluster and NE16), thus showcasing the flexibility of our tool and its management of heterogeneity.
Both HW modules communicate with the control MCU through DMA, fully exploiting double-buffering and interleaving memory transfers with computation.
The cluster and NE16 share an L1 multi-bank memory of 128kB for both activation and weights, while the SoC includes an L2 main memory of 1.5 MB.
The cluster supports a wide range of operators, including pooling, convolutions, additions, and fully-connected layers, all followed by re-quantization. NE16 instead only supports convolutions and fully-connected layers. Note that all patterns present in the NE16 Pattern Table are also included in cluster one, allowing MATCH to choose the best HW Execution Module alternative for these patterns depending on the layer hyperparameters.
Network Transformations for GAP9 have already been shown in Table \ref{tab:net_transformations}. 

Two cost models are defined for the two HW modules: for NE16, we rely on the open-source cost model at\footnote{\url{https://github.com/eml-eda/plinio/blob/main/plinio/cost/ne16_latency.py}}, which allows us to compute the $L_{ops}$ component.
Despite the cluster not being an ``accelerator'' in the proper sense, we still define its \textit{optimal} spatial mapping by considering how the kernels parallelize the innermost loops.
For instance, for a convolution, we set the optimal spatial mapping to be equal to the tile dimensions that maximize the parallelization and memory re-use in the internal loop of the adopted kernel library (PULP-NN), i.e., $OX$ = 2, $K$ = 4, and $OY$ = 8 \cite{garofalo2020pulp}.
When this spatial mapping cannot be used, e.g., when the $OY$ dimension is not multiple of 8, MATCH can select between \textit{padding} and \textit{parallelism reduction}.
For each spatially unrolled dimension (e.g. $OY$), its largest divider smaller than the optimal unrolling factor is used as a new unrolling factor (e.g., $D$, s.t. $OY=nD$, $D<8$). Then, the number of temporal iterations required to process the layer using this new spatial parallelism is computed. If it is identical to the one obtained with the optimal values on a padded version of the input, then the reduced parallelism mapping is kept, as it does not incur a memory overhead. Otherwise, the input is padded. This flexibility is allowed by the structure of the cluster, which does not have a pre-defined spatial unrolling and is supported for all accelerators of this kind in MATCH. Therefore, it is not part of the required target-specific customizations. $L_{ops}$ is finally computed with a model extrapolated from the compiled back-end library, PULP-NN.

For both NE16 and Cluster, since DMA transfers are asynchronous, the total latency is computed as $L = max(L_{ops}, L_{mem,1,2})$, where $L_{ops}$ is modeled as detailed above. $L_{mem,1,2}$ models again the memory transfer cycles and the performance overheads caused by L2-to-L1 DMA calls.
This component is derived from the memory bandwidth information and the transferred tensors dimensions, as in DIANA. In this case, we established a 27-cycles overhead for each chunk of data transferred that is stored contiguously in memory. 

The GAP9 code generator includes APIs for both HW modules; further, it also includes the Synchronization APIs to orchestrate NE16 and Cluster execution, and memory transfers. An example of the API functions for NE16 has been previously shown in Fig. \ref{fig:template}.

%% file: Text/05_Results.tex
\section{Experimental Results}
\noindent
\label{sec:results}
In this section, we evaluate MATCH by compiling different DNNs targeting both DIANA and GAP9.
For both platforms, we set the operating frequency to 260MHz and used the latest available SDKs and open-source kernel libraries.
Latency measurements were obtained using on-board dedicated performance counters.
Our tool is compared with TVM (which only generates code for the main MCU of the two systems) and with HTVM (for DIANA) and DORY (GAP9), which utilize the same platform backend libraries and HW modules but different AI compilation steps. Additionally, on GAP9, we also compare MATCH with the NNTool, a proprietary tool from GreenWaves that utilizes a different backend library.
Noteworthy, all the following results are measured on the actual HW platforms.

In Sec. \ref{sec:micro}, we first report MATCH's results on single convolutional layers, comparing them with plain TVM.
In Sec. \ref{sec:full_network_soa}, we benchmark MATCH and SoA competitors on the four networks from the \textit{MLPerf Tiny} benchmark suite~\cite{mlcommons}: a ResNet V1~\cite{resnet} architecture with a backbone of 8 convolutional layers, trained for image classification on CIFAR10; a MobileNetV1~\cite{howard2017mobilenets} with a width multiplier of $0.25$ for person detection on the  Visual Wake Words dataset; a Depthwise Separable CNN (DS-CNN)~\cite{zhang2017hello} trained for Keyword Spotting (KWS) on the Speech Commands v2 dataset \cite{warden2018speech}, with 105,829 utterances, to be classified into 12 classes; a fully-connected (FC) Autoencoder, which targets the Toy-car data fold included in the DCASE2020 dataset\cite{koizumi2020description}, to detect anomalies based on machine operating sounds.
Finally, in Sec. \ref{sec:ablation}, we perform ablation studies to show the impact of last-level cache memory reduction and demonstrate the performance gain obtained by exploiting SoC heterogeneity.

\subsection{Micro Benchmarking}
\label{sec:micro}
\begin{figure}[t]
    \centering
    \includegraphics[width=\columnwidth]{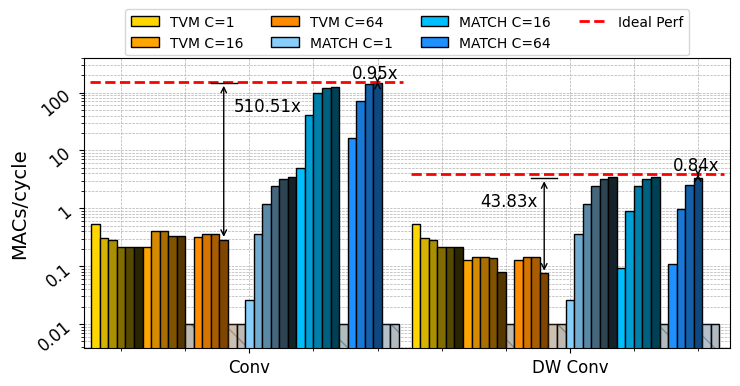}
    \caption{DIANA micro benchmarking results. Darker colors point to bigger spatial dimensions. Grey bars correspond to kernels not fitting in memory.}
    \label{fig:diana_microbench}
\end{figure}
In this section, we present the results of our micro benchmarking experiments. We benchmarked a series of convolutional blocks, encompassing a 2D Conv followed by bias addition and re-quantization (multiplication, right shift, clip, and cast operators). The hyperparameters used were \( IX = IY \in \{2, 8, 16, 32, 64, 128\} \), \( C = K \in \{1, 16, 64\} \), with a padding of 1 for all corners, FX = FY = 3, a convolution stride of 1, and a dilation of 0. We tested classical and depthwise (DW) convolutions with the same hyperparameters.

Fig.~\ref{fig:diana_microbench} showcases the results of MATCH on DIANA's digital accelerator. For standard convolutions, MATCH achieves an average speed-up of 4.43$\times$ compared to TVM with C = 1. As the number of input channels increases, the speed-up also increases, reaching a maximum of 510.51$\times$ for C = 64, IX = IY = 32. With these dimensions, MATCH achieves 146.12 MACs/cycle, which is only 5\% lower compared to the ideal performance predicted by our cost model, demonstrating close-to-optimal compiler management of memory and computation orchestration.
For DW convolutions, the digital accelerator achieves lower efficiency, given its limited spatial utilization. Nevertheless, with C = 64 and IX = IY = 32, MATCH is still able to achieve a speed-up of 43.83$\times$ compared to TVM, achieving 77\% of the estimated MAC/cycle.
Note that even if the accelerator has not been originally designed to execute DW Convolutions, using an appropriate cost model MATCH can still find ways to efficiently execute these layers on it, achieving a significant speed-up.

\begin{figure*}[t]
    \centering
    \includegraphics[width=18cm, height=4cm]{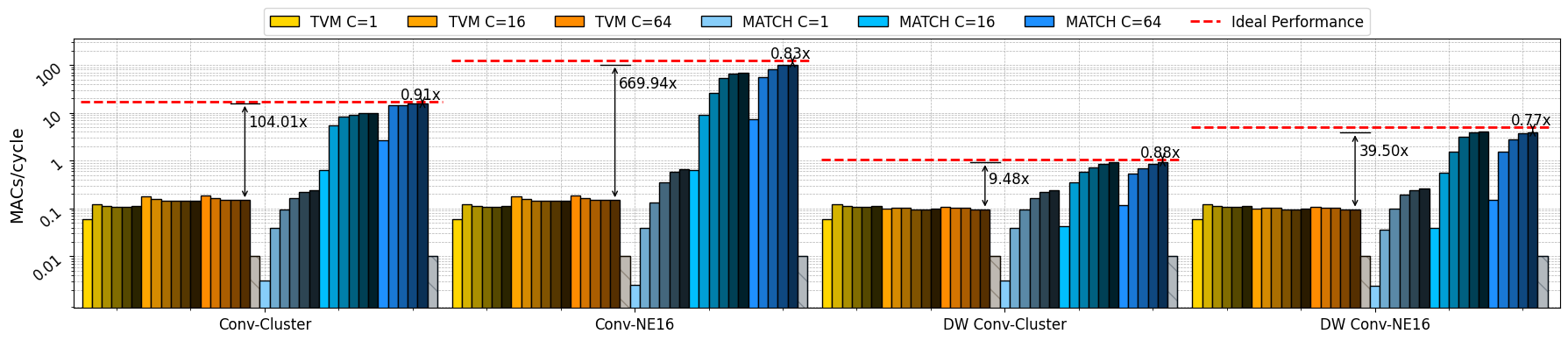}
    \caption{GAP9 Microbenchmarking results. Darker colors point to bigger spatial dimensions. Grey bars point to the kernel not fitting in memory.}
    \label{fig:gap9_microbench}
\end{figure*}
Fig.~\ref{fig:gap9_microbench} presents the results of MATCH on the GAP9 platform, considering both the cluster and the NE16 accelerator. On the cluster, MATCH outperformed TVM by a maximum factor of 104.01$\times$ for standard convolutions and 9.48$\times$ for depthwise convolutions on 64-channels layers.
Noteworthy, these speed-ups do not only derive from the higher number of cores of the cluster compared to the single CPU utilized by TVM: thanks to the backend library that it exploits, MATCH also utilizes SIMDs operations, HW loops, and specific ISA extensions that TVM can not be aware of.
The maximum performance achieved is nearly ideal, reaching 91\% and 88\% of the cost model estimation.
When deployed on the NE16 accelerator, MATCH demonstrated a maximum speed-up of 669.94$\times$ relative to TVM for standard convolutions and 39.5$\times$ for depthwise convolutions. MATCH reached 83\% of the predicted performance for 64-channel convolutions and 77\% for depthwise convolutions, indicating a minor discrepancy between the cost model and actual hardware performance. This suggests that further refinement to the template and specific NE16 backend could enhance the overall final performance. However, the scope of this paper and of our proposed tool is to extract the maximum performance by exploiting a pre-defined existing kernel library and using a universal layer template. We will investigate additional HW-specific optimizations and customizations in our future work.

\subsection{Full Network Execution and Comparison with the State-of-the-Art}
\label{sec:full_network_soa}
\input{tables/full_networks}
In Tab.~\ref{tab:full_networks}, we extend our experiments to the MLPerf Tiny networks on both GAP9 and DIANA, comparing latency results against plain TVM, HTVM, DORY, and NNTool.

When targeting DIANA, MATCH demonstrates significant improvements over TVM, achieving an average speed-up of 60.87$\times$ across the tested networks. Specifically, it achieves 169.4$\times$ lower latency on ResNet, 6.7$\times$ on DSCNN, and 6.4$\times$ on DAE whereas MobileNet cannot be deployed as it exceeds the total on-chip memory.  Similarly, for GAP9, MATCH outperformed TVM substantially, with speed-ups of 47.7$\times$ for MobileNet, 159.2$\times$ for ResNet, 53.1$\times$ for DSCNN, and 11.3$\times$ for DAE.

Compared with HTVM, MATCH exhibits a slight performance deficit for DSCNN and DAE, with latencies of 7.3 ms versus 7.26 ms and 0.4 ms versus 0.36 ms, respectively. This slight disadvantage arises because our layer template is designed to be general across multiple platforms, resulting in a marginally slower execution. However, for MobileNet and ResNet, MATCH improves latency from 6.17 ms and 1.27 ms to 6.08 ms and 0.79 ms, respectively. These gains are attributed to the additional features of MATCH with respect to the tiler included in HTVM, such as the ability to perform loop reordering and uneven tensor mapping, which allows it to find more optimized layer schedules.

When comparing MATCH to DORY, MATCH's ability to leverage the full heterogeneous platform comprising the NE16 accelerator results in an average speed-up across the 4 networks of 2.15$\times$. Still, even when employing just the cluster (precise numbers reported in Tab.~\ref{tab:heterogeneous_impact} below), for MobileNet and ResNet, we achieve latencies of 11.2 ms and 5.48 ms, respectively, outperforming DORY by 1\%/5\%. For DSCNN and DAE, however, MATCH lags behind, with latencies of 4.25 ms and 0.54 ms, due to the layer template generality.

Finally, compared to NNTool, MATCH is, on average, 2.5$\times$ slower. This difference is primarily due to the more efficient and optimized back-end kernels used by NNTool, which are tailored to each specific hyperparameter combination (the library contains hundreds of specialized convolutional kernels), and secondarily to template generalization overheads. Nevertheless, it is important to note that MATCH's framework is designed to be easily extensible, which could in principle allow the integration of alternative back-end kernels, including those used by NNTool.

More in general, MATCH is not designed to provide extremely optimized results on a new platform out-of-the-box, but rather, to support deploying workloads to new accelerators quickly, with few easy customizations.
Overall, the results of this section demonstrate this objective, as MATCH is superior or in the worst case on-par with all other competitors when using the same backend.
While we are outperformed by the ultra-optimized and single-target NNTool, MATCH's modularity features that allow replacing the backend library, and possibly even the DSE engine, could be leveraged to progressively reduce this gap, while maintaining full compatibility with the TVM ecosystem.

\subsection{Ablation study}
\label{sec:ablation}
\begin{figure}[t]
    \centering
    \includegraphics[width=\columnwidth]{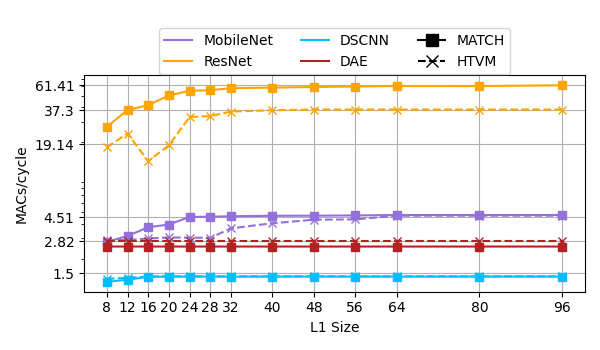}
    \caption{MACs/Cycle of networks deployed on DIANA with scaling L1 size.}
    \label{fig:diana_scaling}
\end{figure}
\begin{figure}[t]
    \centering
    \includegraphics[width=\columnwidth]{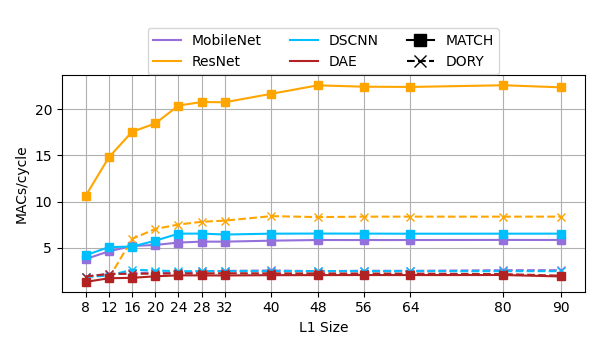}
    \caption{MACs/Cycle of networks deployed on GAP9 with scaling L1 size.}
    \label{fig:gap9_scaling}
\end{figure}
\subsubsection{Scheduling and memory scaling}
In this section, we explore the impact of L1 memory size on the achieved MAC/cycle across the four networks—MobileNet, ResNet, DSCNN, and DAE, on the GAP9 and DIANA hardware platforms (Fig.~\ref{fig:diana_scaling} and Fig.~\ref{fig:gap9_scaling}). The goal is to demonstrate how MATCH enables better scheduling and improved performance with respect to competitors, particularly when memory is constrained.
For the DAE and DSCNN networks, no notable trends are observed across different L1 memory sizes. This is because these networks do not require tiling, allowing for consistent performance regardless of the available memory. The simplicity of these networks allows maintaining relatively stable MAC/cycle performance across varying L1 sizes, for both MATCH and its comparisons.

In contrast, on the MobileNet, at a constrained L1 memory size of 24 kB, on DIANA, we observe that MATCH can still achieve near-full performance, maintaining high MACs per cycle. In this scenario, HTVM's performance drops significantly, falling to 3 MAC/cycle, which results in MATCH achieving a speed-up of 1.5$\times$ over HTVM (versus 1\% at full L1 size). This highlights MATCH's ability to optimize layer scheduling effectively, even with limited memory resources.

For ResNet, the performance of competing tools declines sharply as L1 memory size decreases on both the DIANA and GAP9 platforms. For example, at 16 kB of L1 memory, HTVM becomes 3.01$\times$ slower than MATCH, a stark contrast to the 1.61$\times$ difference observed when using the full L1 size. Similarly, on GAP9, DORY experiences a significant slowdown, being 8.85$\times$ slower than MATCH at 12 kB, compared to a 2.68$\times$ slow-down at full memory capacity. Notably, DORY fails to deploy ResNet at 8 kB of L1 memory on GAP9 because the last layers with high C dimension do not fit, even with the smallest tile size, given the additional buffers that PULP-NN requires. MATCH, on the other hand, successfully deploys ResNet by offloading these layers to the CPU, demonstrating its flexibility and robustness in handling low-memory environments.

\subsubsection{Heterogeneity impact}
\input{tables/heterogeneous_impact}
\begin{figure}[t]
    \centering
    \includegraphics[width=\columnwidth]{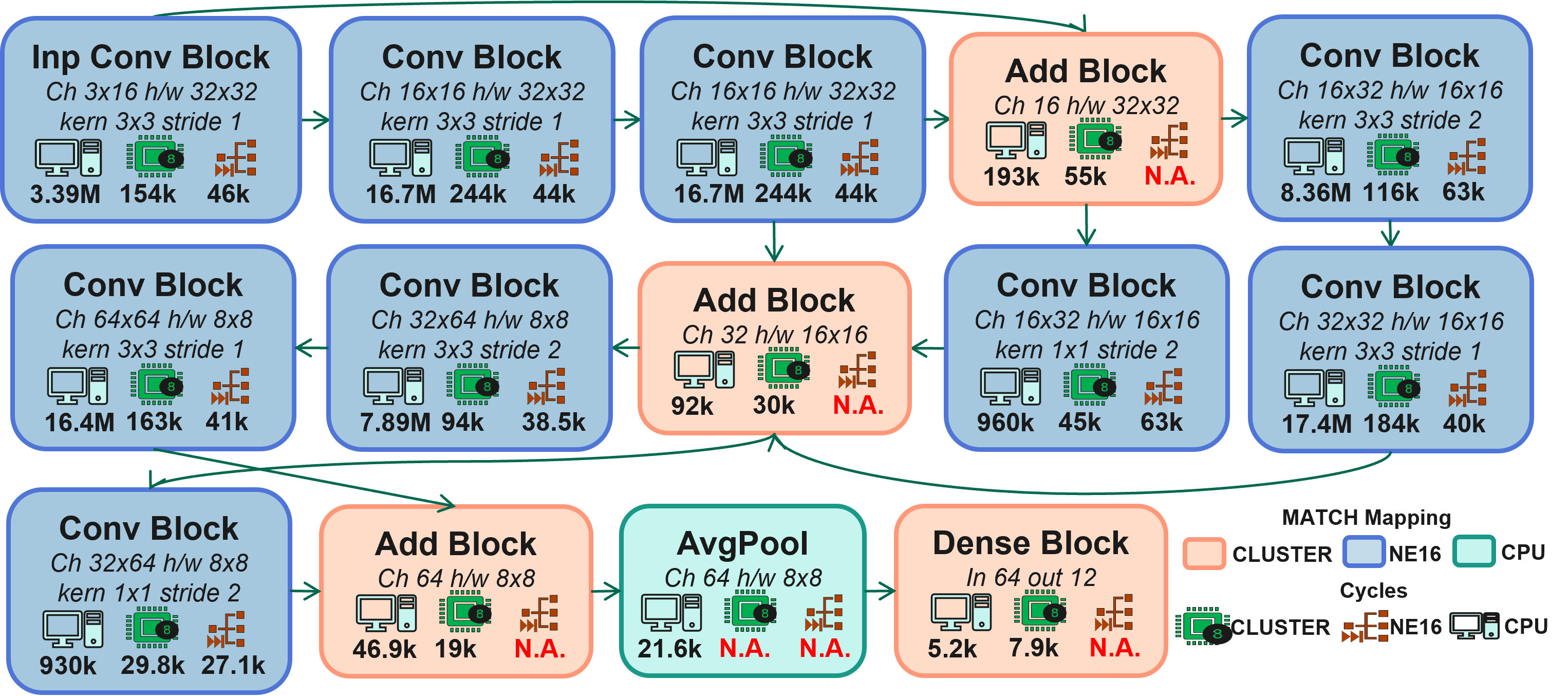}
    \caption{MATCH mapping for the ResNet, displaying for each block the latency, in number of cycles, for each single HW module.}
    \label{fig:gap9_resnet_mapping}
\end{figure}
Tab.~\ref{tab:heterogeneous_impact} summarizes the performance results obtained by enabling different HW modules on the GAP9 \texttt{MatchTarget}. This analysis highlights the flexibility of MATCH in handling multi-accelerator systems, demonstrating the effectiveness of our partitioning mechanism, which groups nodes and assigns them to the HW module that the scheduler predicts will deliver the best performance.

Only enabling the Cluster HW module reduces the latency of the CPU-only solution by an average of 28.64$\times$, with the ResNet network achieving a peak speed-up of 62.58$\times$. Enabling the NE16 HW module without the 8-core cluster achieves an average speed-up of 43$\times$ over the CPU-only baseline, with a maximum speed-up of 118.32$\times$ for the ResNet network.
It is important to note that for the DAE network, the NE16+CPU configuration yields identical results to the CPU-only setup because the DAE model is composed entirely of fully connected layers, which the NE16 library does not support. Similarly, for the DSCNN network, the NE16+CPU configuration results in a lower performance than the Cluster+CPU one since the first layer, featuring a 4x10 rectangular filter, cannot be offloaded to the accelerator. 
Fully leveraging MATCH's heterogeneous deployment capabilities proves to be highly effective, outperforming all other configurations, being on average 67.83$\times$ better than the CPU baseline, 2.13$\times$ better than the Cluster+CPU solution, and 5.7$\times$ better than the NE16+CPU one. 
For the MobileNet network, the multi-target configuration shows a 1.56\% improvement over the NE16+CPU target by offloading the first layer to the cluster, which results in lower latency. In the DSCNN network, the cluster processes the first layer due to the unsupported filter shape by the NE16, while the accelerator efficiently handles the remaining convolutional layers. The full configuration yields results identical to the cluster+CPU setup for the DAE.

In the case of the ResNet model, the full configuration achieves a 34.55\% improvement over the NE16+CPU target and a 2.54$\times$ speed-up over the cluster+CPU configuration. Fig.~\ref{fig:gap9_resnet_mapping} provides a detailed breakdown of the effective latency, in cycles, for each HW module across all ResNet layers, and illustrates MATCH's decision-making process in deciding where to offload each computation. NE16 processes every convolutional layer; the 8-core cluster manages addition operations and the final dense block, while the average pooling operation is handled by the CPU.
Notably, MATCH optimally deploys all layers with only two exceptions. The first is a convolutional layer with parameters \(C = 16\), \(K = 32\), \(OY = OX = 16\), \(FY = FX = 1\), and stride = 2 (second box from the right in the middle row of the figure). This layer is offloaded to the NE16 accelerator, although the 8-core cluster would have been slightly more efficient due to its dimensionality. Once again, this is due to the discrepancy between the estimated and real latency on NE16, caused by inefficiency in the associated kernels. The second exception occurs in the final dense block, where the cluster slightly underperforms compared to the TVM fallback. Although these discrepancies are minor, MATCH’s robustness could be further enhanced by adding more refined cost models and incorporating TVM’s performance metrics into its decision-making process, ensuring that the optimal alternative is always selected.

%% file: tables/full_networks.tex
\begin{table*}[!ht]
    \centering
    \caption{Deployment results for end-to-end networks.\\
    In bold the faster alternative for each platform, for the network, and in italic the second ranked one.\\
    *NNTool is a proprietary source tool which is also platform-specific.}
    \begin{tabular}{c|ccc|cccc}
         & \multicolumn{3}{c|}{\textbf{DIANA}} & \multicolumn{4}{c}{\textbf{GAP9}} \\ \hline
         \textbf{Network} & \textbf{TVM} & \textbf{HTVM} & \textbf{MATCH} & \textbf{TVM} & \textbf{DORY} & \textbf{NNTool*} & \textbf{MATCH} \\ \hline
        \textbf{MobileNet} & OoM & \textit{6.17 ms} & \textbf{6.08 ms} & 236.22 ms & 11.29 ms & \textbf{1.61 ms} & \textit{4.94 ms}\\ 
        \textbf{ResNet} & 133.1 ms & \textit{1.27 ms} & \textbf{0.79 ms} & 342.72 ms & 5.76 ms & \textbf{0.88 ms} & \textit{2.15 ms}\\ 
        \textbf{DSCNN} & 49.16 ms & \textbf{7.26 ms} & \textit{7.3 ms} & 83.41 ms & 4.19 ms & \textbf{0.68 ms} & \textit{1.57 ms}\\ 
        \textbf{DAE} & 2.58 ms & \textbf{0.36 ms} & \textit{0.4 ms} & 6.12 ms & 0.52 ms & \textbf{0.26 ms} & \textit{0.54 ms}\\ \hline
    \end{tabular}
    \label{tab:full_networks}
\end{table*}

%% file: tables/heterogeneous_impact.tex
\begin{table}[!ht]
    \centering
    \caption{GAP9 network deployment using different HW models.}
    \begin{tabular}{ccccc}
    \hline
        \textbf{Network} & \textbf{CPU Only} & \textbf{Cluster+CPU} & \textbf{NE16+CPU} & \textbf{Full} \\ \hline
        \textbf{MobileNet} & 236.22 ms & 11.2 ms & 5.02 ms & 4.94 ms \\ 
        \textbf{ResNet} & 342.72 ms & 5.48 ms & 2.9 ms & 2.15 ms \\ 
        \textbf{DSCNN} & 83.41 ms & 4.25 ms & 14.46 ms & 1.57 ms \\ 
        \textbf{DAE} & 6.12 ms & 0.54 ms & 6.12 ms & 0.54 ms \\ \hline
    \end{tabular}
    \label{tab:heterogeneous_impact}
\end{table}

%% file: Text/06_Conclusion.tex
\section{Conclusions}
\label{sec:conclusion}
\noindent We introduced MATCH, a novel model-aware toolchain for efficient DNN deployment on accelerators.
Differently from other target-specific toolchains, MATCH does not embed hardware-dependent optimizations or heuristics in the code but rather exposes an API to define high-level model-based hardware abstractions, fed to a generic and flexible optimization engine.
As a consequence, adding support for a new HW module becomes significantly easier, avoiding complex optimization pass re-implementations. A new HW target can be added in less than 1 week of work.
Besides generalization capability, with experiments on several deep learning workloads and end-to-end networks, we have shown that MATCH can achieve performances comparable to highly optimized and HW-specific toolchains on the DIANA platform and on GAP9, which includes two very different HW modules (the 8-core cluster and the NE16 accelerator), proving the tools' flexibility and the capability of handling heterogeneous targets. 
Compared to the SoA open-source solution, on the four MLPerf Tiny benchmarks, we obtained 16.94\% lower latency on DIANA and 2.15$\times$ lower latency on GAP9.

%% file: Text/Acknowledgments.tex
\section{Acknowledgments}
\noindent This work has received funding from the Key Digital Technologies Joint Undertaking (KDT-JU) under grant agreement No 101095947 and grant agreement No 101112274. The JU receives support from the European Union’s Horizon Europe research and innovation program.
The authors acknowledge the support of the Italian Ministry of University and Research (MUR) and the Sustainable Mobility Center (MOST) through the project PNRR - M4C2 - CNMS - Spoke 2, funded under the scheme CN00000023 - PNRR – M4C2 Inv. 1.4 with grant agreement no. 55\_PRR22\_1112\_22\_AT002129.